\documentclass[%
  onecolumn
   , colorlinks % also like hyperref with CSC colorscheme (this is the default)
]{mpi2015-cscpreprint}

\usepackage[american]{babel}

\usepackage{graphicx}
\usepackage{hyperref}
\usepackage{tikz}
\usetikzlibrary{calc}
\definecolor{mardiorange}{HTML}{D0662B}
\definecolor{mardiblue}{HTML}{005EAA}
\colorlet{JHcolor}{mardiblue}

\def\mf{\texttt{MaRDIFlow}}

\usepackage{amssymb}
\usepackage{amsthm}

\usepackage[vlined,longend,linesnumbered,ruled]{algorithm2e}

\begin{document}
  
\title{MaRDIFlow: A CSE workflow framework for abstracting meta-data from FAIR computational experiments}
  
\author[$\ast$]{Pavan L. Veluvali}
\affil[$\ast$]{Max Planck Institute for Dynamics of Complex Technical Systems, Sandtorstr.~1, 39106 Magdeburg\authorcr
  \email{\{veluvali, heiland, benner\}@mpi-magdeburg.mpg.de}; \orcid{0000-0001-8804-0338}, \orcid{0000-0003-0228-8522}, \orcid{0000-0003-3362-4103}}
  
\author[$\ast$]{Jan Heiland}
  
\author[$\ast$]{Peter Benner}
    
\shorttitle{}
\shortauthor{PL. Veluvali, J. Heiland, P. Benner}
\shortdate{}
  

\abstract{%
	Numerical algorithms and computational tools are instrumental in navigating and
addressing complex simulation and data processing tasks. The exponential growth 
of metadata and parameter-driven simulations has led to an increasing 
demand for automated workflows that can replicate computational experiments across platforms. In general, a
computational workflow is defined as a sequential description for
accomplishing a scientific objective, often described by tasks and their associated data
dependencies. If characterized through input-output relation, workflow components can be structured to allow interchangeable utilization of individual tasks and their accompanying metadata. In the present work, we develop a novel computational framework, namely, \mf, that focuses on the automation of abstracting meta-data embedded in an ontology of mathematical objects. This framework also effectively addresses the inherent execution and environmental dependencies by incorporating them into multi-layered descriptions.
Additionally, we demonstrate a working prototype with example use cases and methodically integrate them into our workflow tool and data provenance framework. Furthermore, we show how to best apply the FAIR principles to computational workflows, such that abstracted components are Findable, Accessible, Interoperable, and Reusable in nature. }

\novelty{This manuscript presents a novel research data management tool, \mf~, that focuses on the automation of abstracting meta-data embedded in an ontology of mathematical objects. The working prototype of this software tool is illustrated through FAIR computational experiments.}

\maketitle

\section{Introduction}

The interplay of data-intensive computational studies is a substantial part of
scientific endeavors across all disciplines. 
Computational workflows have been used as a systematic way of describing the
methods needed, the data involved, as well as computing resources and infrastructures. 
With ever more complex simulation models and ever larger primary data
volumes, CSE (Computational Sciences and Engineering) workflow descriptions
themselves have become
an enabler for research beyond the execution of simulations, for example,
to extract latent information from various data repositories and compare methodologies across diverse data and computational frameworks \cite{atkinson2017scientific}.

The FAIR principles \cite{wilkinson2016fair}, describe a set of requirements for data management and stewardship to ensure that the research data are Findable, Accessible, Interoperable, and Reusable. Each guiding principle is proposed to define the degree of `FAIRness' via describing the distinct considerations for contemporary environments, tools, vocabularies and data infrastructures. While the elements of FAIR Principles are related and separable, they are equally applied to identify, describe, discover, and reuse meta-data assets of scholarly outputs. Overall, FAIR principles act as a guide to assist data stewards in evaluating their implementation choices. More recently, they have been adopted by funding agencies, such as the German Research Foundation \cite{dfg} for developing assessment metrics of research metadata across various disciplines \cite{devaraju_anusuriya}.

While there is a knowledge base for CSE workflows from a (software) engineering point of view \cite{Her09, BriCGetal14} and while it has been acknowledged that for documentation, model descriptions and code can complement each other \cite{FehHHS16}, an inclusive abstract description of CSE workflows is not yet 
anchored. As for combining models, code, and data for the description of CSE simulations in a virtual lab notebook, Jupyter notebooks have gained popularity \cite{KluRPetal16}. Also services like Code Ocean \cite{clyburne2019computational}   
target the combination of code and model descriptions. Still, little effort has been made to use abstraction for CSE workflow components in view of documentation tools that are generally applicable and that scale well with ever more demanding and sophisticated simulations. Lately, with the advancement of data intensive research, there has been a rise in the development of automated and reusable workflows, wherein these workflows aim to seamlessly integrate computer-based and laboratory computations through artificial intelligence \cite{national2022automated}. 

In this work, we analyze general and particular components and provide an abstract multi-layered description of CSE workflows: Each component will be characterized through an input/output description so that model, data, and code can be used interchangeably and, in the best case, redundantly. For that, we describe suitable meta data and a low level language for the descriptions of general CSE workflows \cite{veluvali2023mardiflow}. Additionally, we emphasize that the introduction of redundancy in the representation of models, code, and data serves as a positive feature for CSE workflows. This redundancy enhances the robustness of workflows via ensuring compatibility during potential execution issues. With interchangeable and  multi-level components, workflows become more adaptable and reproducible, contributing to the overall reliability of a scientific task.
Generally, we understand a CSE workflow as a chain of one or more interconnected models used for simulations. From existing literature, a CSE workflow is defined as a precise description of a multi-step procedure to coordinate multiple tasks and their meta-data dependencies \cite{goble2020fair}. In workflow systems, each task is represented through the execution of a computational process, such as, executing a code, calling a command line tool, accessing a database, submitting a job to a HPC cluster, or executing a data processing script.

In their general treatment, the following constraints need to be taken into account
\begin{enumerate}
  \item In particular in a CSE context, each model might be arbitrarily complex and computationally demanding. 
  \item Often, the particular numerical realization represents a compromise between accuracy and computational costs.
  \item Within a workflow, models are likely implemented in different frameworks or languages.
  \item In the case of, say, commercial codes that may well be one part of a workflow, some simulation models might not be fully available but only evaluable through interfaces.
  \item Possibly, the actual simulation code is not available at all but only descriptions and, in the better case, alternative implementations.
\end{enumerate}

Nevertheless, the goal of any CSE workflow framework is to offer a specialized programming environment that minimizes the efforts required by scientists or researchers to perform a computational experiment \cite{veluvali2023mardiflow}. In general, CSE workflow description can be categorized into distinct parts or phases, as listed below, governing its functional operation.

\begin{itemize}
\item Composition and abstraction
\item Execution
\item Meta-data mapping and provenance
\end{itemize}

Firstly, during composition and abstraction, a CSE workflow is created either from scratch or from modifying a previously designed workflow, whereby the user relies on different workflow components and data catalogs. Some of the well-known methods for editing and composing workflows are either textual or graphical or mechanism-based semantic models \cite{deelman2009workflows}. The workflow then abstracts software components written by third parties, and handles heterogeneity via shielding run time incompatibility and complexities. Secondly, during execution, the workflow components are executed either by a computational engine or via a subsystem, wherein a static or an adaptive model is implemented to realize the meta-data. Importantly, repetitive and reproducible pipelines in order to manage the control and flow of a simulation are an important aspect of the second phase. Once the workflow is well defined, all, or portions of the workflow are sent for mapping. Finally, the data and all associated metadata and provenance information are recorded and placed in user-defined registries which are then accessed to design a new workflow description. Through the following stages, CSE workflow description act as modular building blocks with standardized interfaces, and are generally linked and run together by a computational framework. 

The present article is organized as follows: in the following section, we discuss the current state of the art in Jupyter notebooks and computational workflows. Afterwards, we present our research data management tool, namely, \mf, wherein the framework and its usage as a command-line tool is discussed in detail. Next, we discuss our RDM tool via minimum working examples. Lastly, we put forward the conclusions and future direction from the present work.

\section*{Existing Solutions}
Reproducibility poses a multifaceted challenge which often demands a comprehensive examination. In this work, we focus on the specific aspect of ensuring reproducibility of computational workflows through integrated software solutions.  

Because of the popularity and universality of Jupyter notebooks, we begin with
highlighting the capabilities of the Jupyter environment, which significantly boost productivity in computational science and mathematics, while also promoting reproducibility \cite{beg2021using}.
In general, Jupyter Notebooks \cite{jupyter} are accessed through a modern web browser and are typically designed to support interactive exploration and publishing records of a scientific computation. Through text and code blocks, it performs a specific computation and elucidates it in detail. The code within a given Jupyter Notebook is organized into cells, which in turn allows individuals to modify and execute, respectively. Also, the output from each cell appears directly below it and is stored as part of the document. This approach often facilitates a symbiotic display of code, data, and model descriptions and naturally ensures replicability of the experiments; \cite{FehHHS16}. In this respect, the side by side appearance of text (for documentation) and code (for the execution) is in line with the redundant or multi-layered representation of workflows that we want to achieve but adds to the complexity of such a notebook realization. 

What concerns reproducibility, in the strict sense that an experiment could be reproduced solely by information provided in the notebook, certain design decisions in the Jupyter notebook like undocumented versions of the imported libraries let alone the underlying libraries in the backend may stand in the way. From a more practical perspective, by its linear design of consecutive cells, Jupyter notebooks are not well suited to handle larger projects. And although a call of third party code is certainly possible through direct Julia/python/R interfaces or through the system's shell, the embedding of external tools is not a primary and, thus, not a well-defined feature of Jupyter notebooks. Finally, the reproducibility of a workflow in a Jupyter notebook hinges on the code only whereas the description is commonly seen as an add-on. Thus, there is no built-in mechanism that ensures completeness of the documentation to function as an equivalent or fully valid addition to the code base.
In fact, recent studies have found that a Jupyter notebook per
  se is not a strong guarantee for reproducibility; see Refs. \cite{SamM24,PimMBF21}.
  
  Whereas Jupyter notebooks have become a popular tool for implementing,
documenting, and publishing workflows and experiments of moderate complexity, in
computational science and engineering, multiple advanced tools have addressed
the challenge of managing workflows composed of various simulation codes. Domain
specific workflow managers, such as CWL \cite{cwl} and Galaxy \cite{galaxy} are
applicable to high-performance computing in general. CWL \cite{cwl} is widely
known for the description of command-line tools and of the workflows made from
these tools. It includes many features, such as, software containers, resource
requirements, workflow-level conditional branching. While most of the CWL
development began in the field of bioinformatics, since 2016, the CWL standards have been used in other fields, including hydrology, radio astronomy, geo-spatial analysis, and high-energy physics. On the other hand, Galaxy \cite{galaxy} serves as an accessible browser-based platform for scientific computing. It facilitates data sharing, analysis, and visualization for scientists with minimal technical barriers. In addition, integration with third-party tools like \emph{noWorkflow} \cite{pimentel2017noworkflow} allows for effective tracking of provenance, elucidating the relationship between inputs, code, and generated files.

A container-based approach to modelling workflow components has been followed in the \emph{functional mockup interface} (FMI) \cite{blochwitz2011functional} development. Here, arbitrary simulation model are encapsulated in a complete virtual computational environment (a container) and made accessible through interfaces. This facilitates an easy exchange of simulation tools (even without disclosing the source) and can be integrated in workflow designs, as it is specified in the FMI using \emph{xml} syntax.

Most of the aforementioned systems have improved the reproducibility of
computational workflows over the last years, and have become the defacto
standard for syntactic interoperability of workflow management systems. However,
each of the systems discussed come with their own limitations. Namely, CWL often syntax fails to address the user-defined construction and interaction with other command-line tools once its execution is finished \cite{crusoe2022methods}. Likewise, one of the disadvantages of document based workflow definitions is their static nature, as the exact flow of the workflow must be known before execution. This specific mechanism inherently imposes constraints on programming structures that can be utilized, generally via confining options to either directed acyclic graphs (DAGs) or directed cyclic graphs (DCGs), especially in cases where loops are accommodated by the markup language \cite{uhrin2021workflows}. Nevertheless, all the aforementioned frameworks store the data produced, but none with a focus on explicitly recording provenance in detail, and, in particular, the workflow components are not expressed in a multi-level framework or abstract objects. On the other hand, customizing the system for specific requirements may present a significant challenge for end users, requiring additional effort and resources.

Container-based implementations like in FMI can mitigate the issues with data provenance or setting up the environment as much as mere replication is concerned. However, in view reproducibility in different environments or reusability and adaptation of workflow components, these containers would need to be equipped with just the same meta-data as any other workflow design tool would require it.

Based on the aforementioned challenges, it becomes evident that the existing research landscape for a comprehensive workflow description tailored to effectively handle mathematical data within a multi-component framework is lacking. On that account, the present study aims to bridge the existing gap.

\section{MaRDIFlow}

\begin{figure}
	
	\newlength\boxheight
	\setlength{\boxheight}{0.2\linewidth}
	
	\begin{center}
		Level 1: Mathematical Model\\[.05in]
		\def\bone{Electromagnetics}
		\def\btwo{Mechanics}
		\def\bthree{Acoustics}
		\begin{tikzpicture}

\tikzset{rectangle/.append style={line width=2pt, fill=white,
         % rounded corners=2pt,
         minimum height=.5\boxheight}}

% %% center box
\node[rectangle, name=mechanics, draw=JHcolor,
      % xshift=-2\boxheight,
      align=center, text width=\boxheight]
  {{\btwo}}
  ;

\node[rectangle, draw=JHcolor, name=acoustics, xshift=1.25\boxheight,
      align=center, text width=\boxheight]
  at (mechanics.east)
  {{\bthree}};

\node[rectangle, draw=JHcolor, name=electro, xshift=-1.25\boxheight,
      align=center, text width=1.3\boxheight]
  at (mechanics.west)
  {\bone};

\draw[<->,line width=2pt,JHcolor] 
  let
    \p1=(electro.east),
    \p2=(mechanics.west)
  in
  (\x1,\y1) -- (\x2,\y2);

\draw[<->,line width=2pt,JHcolor] 
  let
    \p1=(mechanics.east),
    \p2=(acoustics.west)
  in
  (\x1,\y1) -- (\x2,\y2);

\end{tikzpicture}\\[.1in]
		Level 2: Simulation Model\\[.05in]
		\def\bone{FE-Model 1} \def\btwo{FE-Model 2} \def\bthree{FE-Model 3}
		\begin{tikzpicture}

\tikzset{rectangle/.append style={line width=2pt, fill=white,
         % rounded corners=2pt,
         minimum height=.5\boxheight}}

% %% center box
\node[rectangle, name=mechanics, draw=JHcolor,
      % xshift=-2\boxheight,
      align=center, text width=\boxheight]
  {{\btwo}}
  ;

\node[rectangle, draw=JHcolor, name=acoustics, xshift=1.25\boxheight,
      align=center, text width=\boxheight]
  at (mechanics.east)
  {{\bthree}};

\node[rectangle, draw=JHcolor, name=electro, xshift=-1.25\boxheight,
      align=center, text width=1.3\boxheight]
  at (mechanics.west)
  {\bone};

\draw[<->,line width=2pt,JHcolor] 
  let
    \p1=(electro.east),
    \p2=(mechanics.west)
  in
  (\x1,\y1) -- (\x2,\y2);

\draw[<->,line width=2pt,JHcolor] 
  let
    \p1=(mechanics.east),
    \p2=(acoustics.west)
  in
  (\x1,\y1) -- (\x2,\y2);

\end{tikzpicture}\\[.1in]
		Level 3: Surrogate Model\\[.05in]
		\def\bone{Data Base} \def\btwo{Neural Network} \def\bthree{ROM}
		\begin{tikzpicture}

\tikzset{rectangle/.append style={line width=2pt, fill=white,
         % rounded corners=2pt,
         minimum height=.5\boxheight}}

% %% center box
\node[rectangle, name=mechanics, draw=JHcolor,
      % xshift=-2\boxheight,
      align=center, text width=\boxheight]
  {{\btwo}}
  ;

\node[rectangle, draw=JHcolor, name=acoustics, xshift=1.25\boxheight,
      align=center, text width=\boxheight]
  at (mechanics.east)
  {{\bthree}};

\node[rectangle, draw=JHcolor, name=electro, xshift=-1.25\boxheight,
      align=center, text width=1.3\boxheight]
  at (mechanics.west)
  {\bone};

\draw[<->,line width=2pt,JHcolor] 
  let
    \p1=(electro.east),
    \p2=(mechanics.west)
  in
  (\x1,\y1) -- (\x2,\y2);

\draw[<->,line width=2pt,JHcolor] 
  let
    \p1=(mechanics.east),
    \p2=(acoustics.west)
  in
  (\x1,\y1) -- (\x2,\y2);

\end{tikzpicture}
	\end{center}
	\caption{Generic chain of models that describe the workflow in the simulation
		of transformer noise generation and its realization on different levels of
		abstraction.}
	\label{fig:generic-workflow}
\end{figure}
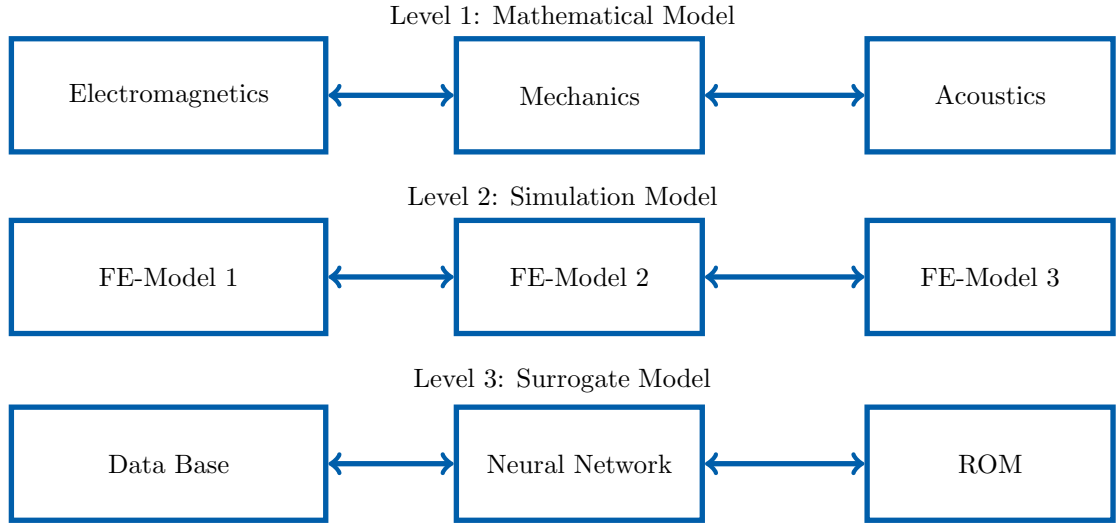

The design principle of \mf~is to handle the components as abstract objects described by their input to output behavior and metadata. By means of the metadata and by matching the I/O interfaces, the objects can be chained together to form a workflow; see Fig. \ref{fig:generic-workflow}.

Each item then can have different descriptions or realizations that are, in the best case, equivalent and redundant; see Fig. \ref{fig:generic-wf-vertical} for an example of such a vertical dimension in the workflow. We note that the redundancy is meant from a theoretical perspective. Practically, the different realizations or representations can be used in different scenarios like compiling a mathematical description of the workflow or using lookup tables as a shortcut during a simulation.

Additionally, this multi-level description enhances or even enables reproducibility in scenarios where, for example, certain software components of a workflow are not available but will be replaced by data or a description and vice versa.

\begin{itemize}
	\item Defined by metadata
	\item Realized as Fig.~\ref{fig:generic-workflow} and Fig.~\ref{fig:generic-wf-vertical}
\end{itemize}

\setlength{\boxheight}{0.2\linewidth}
\colorlet{JHcolor}{mardiblue}
\begin{figure}
	\def\bone{Mathematical Model}
	\def\btwo{Finite Element Model}
	\def\bthree{Reduced Order Model}
	\begin{center}
		\begin{tikzpicture}

\tikzset{rectangle/.append style={line width=2pt, fill=white,
         % rounded corners=2pt,
         minimum height=.5\boxheight}}

% %% center box
\node[rectangle, name=mechanics, draw=JHcolor,
      % xshift=-2\boxheight,
      align=center, text width=1.3\boxheight]
  {{\btwo}}
  ;

\node[rectangle, draw=JHcolor, name=acoustics, yshift=-.5\boxheight,
      align=center, text width=1.3\boxheight]
  at (mechanics.south)
  {{\bthree}};

\node[rectangle, draw=JHcolor, name=electro, yshift=.5\boxheight,
      align=center, text width=1.3\boxheight]
  at (mechanics.north)
  {\bone};

\node[rectangle, draw=JHcolor, name=woe, yshift=-.5\boxheight,
      align=center, text width=1.3\boxheight]
  at (acoustics.south)
  {{I/O Data}};

\node[name=woeee, yshift=-.12\boxheight, fill=white,
      align=center]
  at (electro.south)
  {{... or ...}};

\node[name=woeee, yshift=-.12\boxheight, fill=white,
      align=center]
  at (mechanics.south)
  {{... or ...}};

\node[name=woee, yshift=-.12\boxheight, fill=white,
      align=center]
  at (acoustics.south)
  {{... or ...}};

% \draw[<->,line width=2pt,JHcolor] 
%   let
%     \p1=(electro.south),
%     \p2=(mechanics.north)
%   in
%   (\x1,\y1) -- (\x2,\y2);
% 
% \draw[<->,line width=2pt,JHcolor] 
%   let
%     \p1=(mechanics.south),
%     \p2=(acoustics.north)
%   in
%   (\x1,\y1) -- (\x2,\y2);

\end{tikzpicture}
	\end{center}
	\caption{An exemplified vertical multi-level dimension of a MaRDIflow
		component: equivalent and preferably redundant descriptions of a workflow
		unit.}
	\label{fig:generic-wf-vertical}
\end{figure}
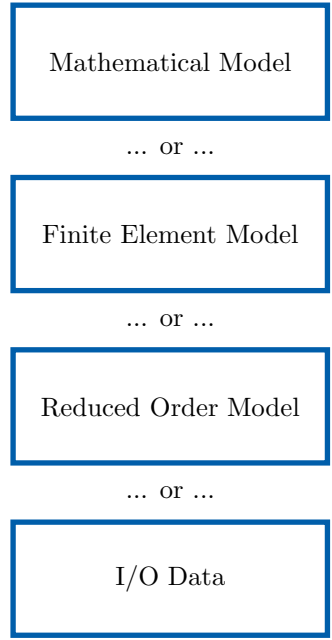

\begin{figure}[th!]
	\centering
	\includegraphics[width=0.6\textwidth]{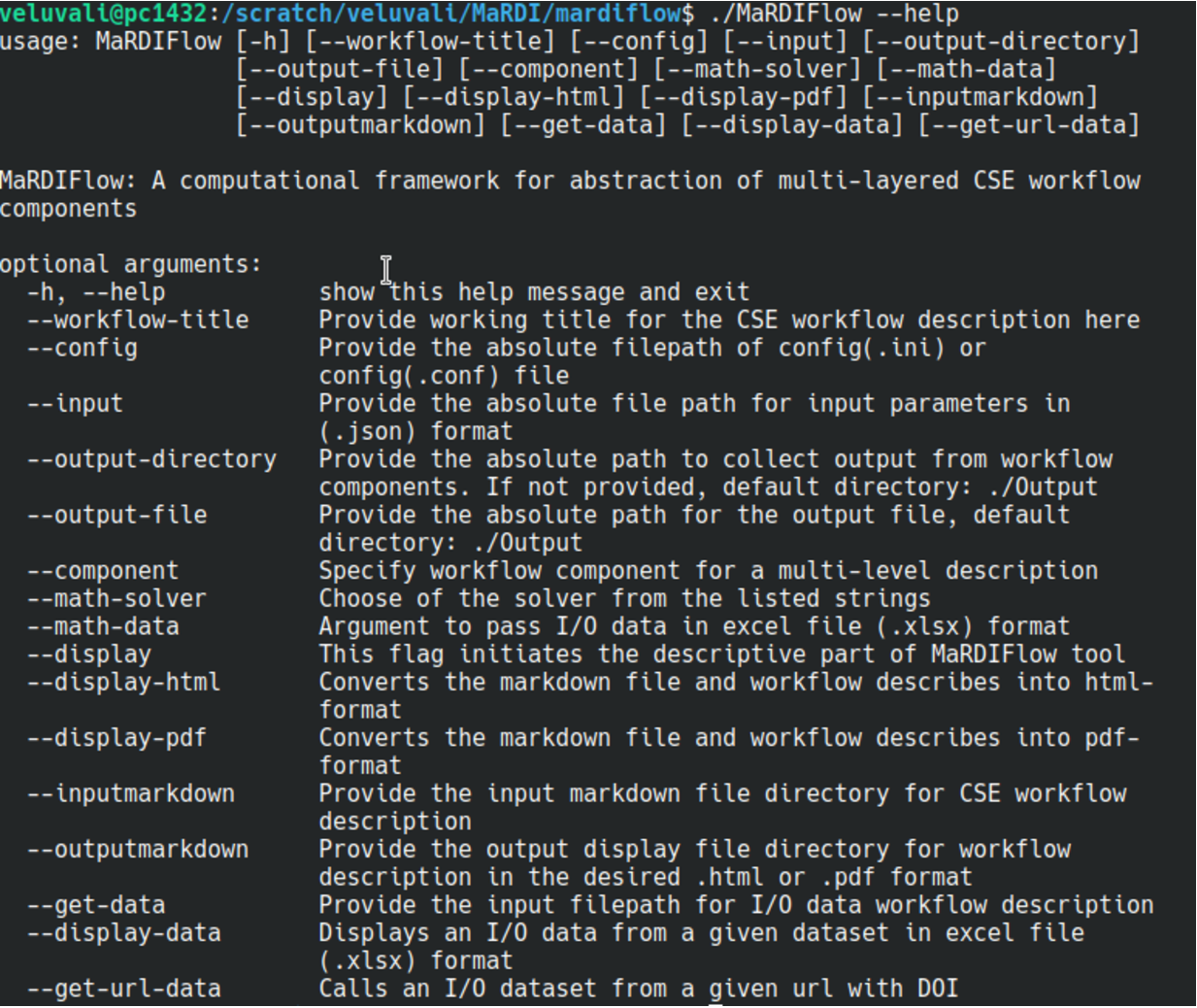}
	\caption{A screenshot illustrating the help message of our RDM tool, \mf.}
	\label{console}
\end{figure}

The working prototype of \mf~is designed as a command line tool that enables researchers and users to execute, document and maintain the provenance for reproduction and replication of computer-based experiments. As shown in Fig.~\ref{console} via a screenshot, \mf~\texttt{--help} option provides a help message with an immediate sense of what MaRDIFlow is. Through the following list we discuss some of the important arguments required to configure and perform the most common tasks with our RDM tool. 

\begin{figure}[th!]
	\centering
	\includegraphics[width=0.4\textwidth]{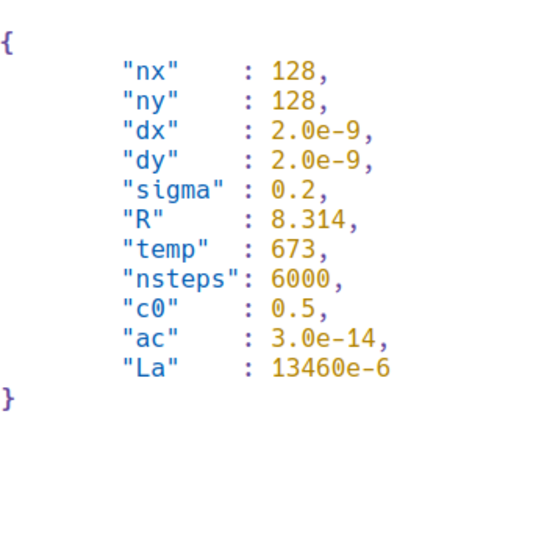}
	\caption{Example inputs object JSON file format with a set of static parameters required for a specific workflow component.}
	\label{json_input}
\end{figure}

\begin{figure}[th!]
	\centering
	\includegraphics[width=0.65\textwidth]{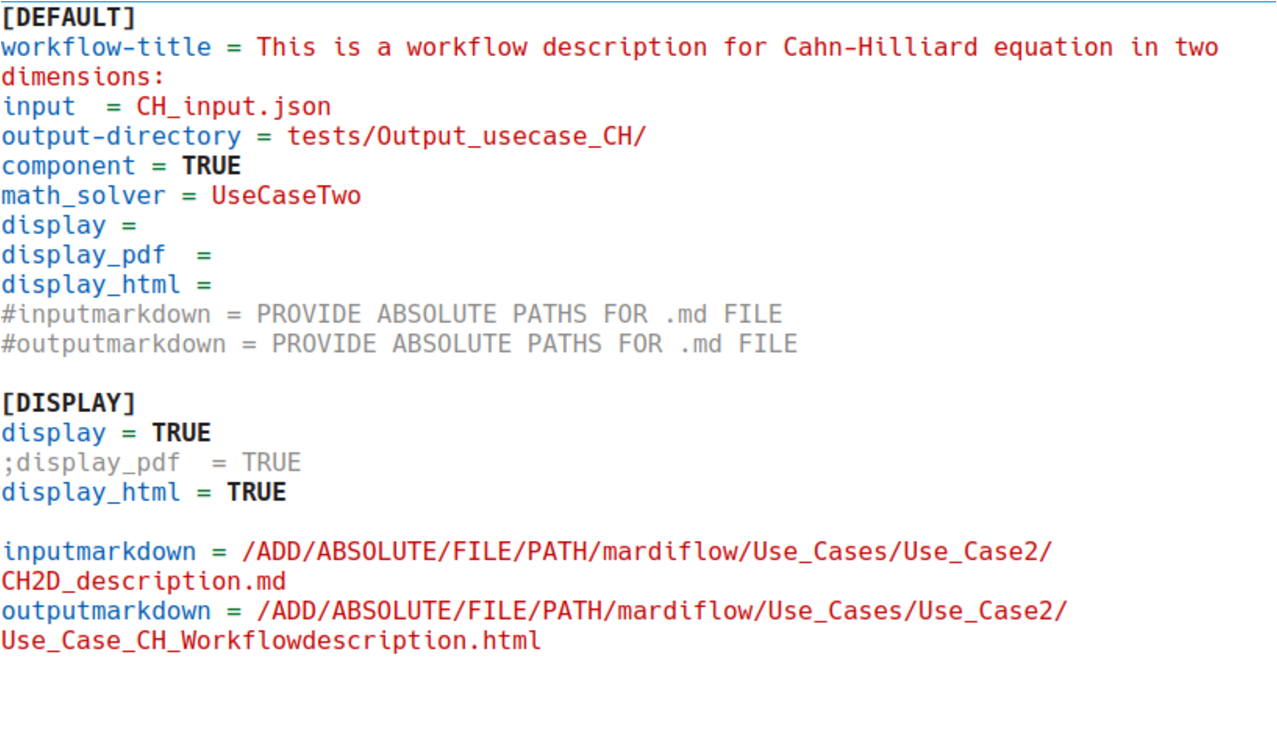}
	\caption{Screenshot of an example configparser .ini file to initialize and run \mf~through command-line.}
	\label{config_file}
\end{figure}

\begin{itemize}
	\item \texttt{--workflow-title} provides the working title for the workflow, as default we have, \texttt{workflow\_title = This is a CSE workflow description under MaRDIFlow}.
	
	\item \texttt{--input} requires an inputs object file in \texttt{.json} format such that it consists of all the numerical parameters for the workflow. An example inputs object file is shown in Fig.~\ref{json_input}, where the key value pairs are a valid string representing the parameter name, and the values found to the right side of the colon are the absolute values for the given input parameter. As shown in Fig.~\ref{json_input}, the JSON object represents the simulation and material parameters which are then passed on to the required workflow component.
	
	\item \texttt{--output-directory} argument allows the user to specify the desired output directory. If not provided, then the workflow output is collected in the root working directory, namely \texttt{Output}. 
	
	\item \mf~can also be configured using a \texttt{.ini} config parser file, which includes both the required default and user-defined sections. It also acts as a list of parameters that governs how the tool is run and configured.
	
	\item \texttt{--config}~\texttt{example\_config\_file.ini} will execute \mf~through terminal. An example configparser file is showcased via a screenshot in Fig.~\ref{config_file}.
	
	\item \texttt{--component} argument is necessary to execute the desired workflow component in the multi-level framework, as previously discussed. In present version, users can choose between \texttt{--math-data} or \texttt{--math-solver} components for executing a given I/O data or a numerical model, respectively. 
	
	\item \texttt{--display} represents the descriptive part of our RDM tool. Herein, \texttt{--display\_html=TRUE} or \texttt{--display\_pdf=TRUE} will convert the \texttt{.md} into desired format. An empty string will parse \texttt{FALSE} bool value. The file path for workflow description in \texttt{.md} file format is passed on to the workflow tool via \texttt{--inputmarkdown} flag. For this flag, it is important that the absolute file path for input as well as for output is provided.
	
	\item \texttt{--data} configures the second component of the workflow description for a given I/O data. By utilizing \texttt{--get-data} and \texttt{--get-url-data} flags, users can furnish a workflow description for a lookup table or a database, thereby presenting an alternative approach to a numerical model. 
	
\end{itemize}

Overall, the working prototype of \mf~ideally provides the user with a specification description, that elucidates the meta-data for a given use case. In addition, it also consists of a computational part that acts as a mechanism to call and execute the given meta-data. In order to further understand our RDM tool, the following minimum working examples implemented in ~\mf~are discussed below. 

\section*{Minimum working examples}
In the present section, we present some use cases as minimum working examples to illustrate in detail the working prototype of ~\mf. 

\subsection*{Methanization Reactor}
\begin{figure*}[th!]
\centering
\includegraphics[width=1.0\textwidth]{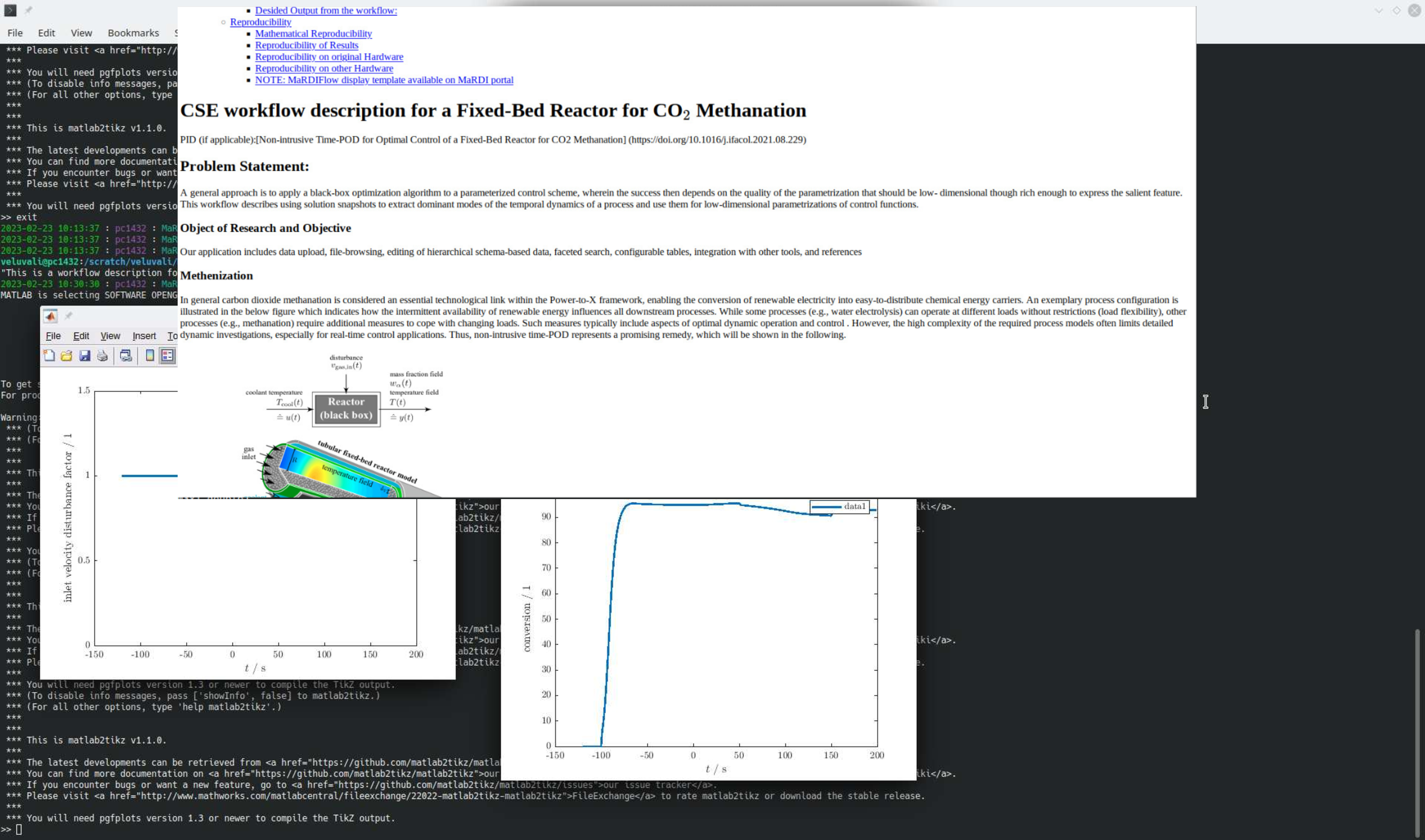}
\caption{A screenshot illustrating the workflow description of a methanization reactor model using the \mf~tool. As an output, screenshots of the console and trajectory of the changing reactor load are shown here from the workflow. In addition, as a final output, the descriptive part of the tool shows the necessary meta-data to reproduce the use case.}
\label{MRworkflow}
\end{figure*}

As a working example within the \mf~framework, a forward solution that converts $\textrm{CO}_{2}$ to $\textrm{CH}_{4}$ as a result of methanization \cite{bremer2021non} is illustrated. In general, reactor models are crucial for converting renewable electricity into chemical energy carriers, specifically through carbon dioxide methanation \cite{bremer2021non}. In this study, a reactor model was examined through a set of nonlinear partial differential equations (PDEs) for mass and energy balances.  The schematic representation of the workflow is provided in Fig.~\ref{MRworkflow}, and we write the governing PDEs as

\begin{equation}
\epsilon\rho_{gas}\frac{\partial{w}_\alpha}{\partial{t}} = -\rho_{gas}\mathbf{v}\,\nabla\cdot{w_{\alpha}} - \nabla\cdot\mathbf{{j_{\alpha}}} + (1-\epsilon)\nu_{\alpha}{\widetilde{r}}
\end{equation}

\begin{equation}
(\rho{c_{p}})_{eff}\frac{\partial{T}}{\partial{t}} = -(\rho{c_{p}})_{gas}\mathbf{v}\cdot\nabla{T} - \nabla\cdot{\dot{{\mathbf{q}}}} + (1-\epsilon)\Delta_{R}{\tilde{H}}\widetilde{r}
\end{equation}

In the above set of equations, $w_{\alpha}$ is the component of mass fraction for 
$\alpha \in{\textrm{CO}_{2}},\textrm{H}_{2}\textrm{CH}_{4},\text{H}_{2},\textrm{O}$, $\mathbf{v}$ is the superficial gas velocity, $\rho$ is the gas density, $\mathbf{j}_{\alpha}$ are dispersive component fluxes, $\nu_{\alpha}$ is stoichiometric coefficients, $\widetilde{r}$ is the reaction rate of methanation, $c_p$ is the specific heat, $T$ is the temperature, $\mathbf{\dot{q}}$ and $\Delta_R\tilde{H}$ is the heat reaction. Furthermore, this study revolves around to ensure that the reactor operates at maximum conversion rates, even when subjected to varying loads. To address this specific objective, an optimal control problem (OCP) is defined as follows:

\begin{equation}
\zeta(u) = \frac{1}{t_e}\int_{0}^{t_{e}} X_{CO_{2}}(t) \,d{t}
\end{equation}

Here, we express the reactor model as a system of ordinary differential equations (ODEs), obtained through the finite volume method applied to the PDE system. Additionally, on states and control we incorporate inequality constraints. Further initialization and model specific details can be found in Ref. \cite{bremer2021non}. Workflow description for the methanization reactor model is defined as given below:

\begin{itemize}
    \item Initialize the system with the given set of governing equations
    \item Perform a forward simulation with temperature as the input parameter
    \item Calculate the conversion rates via calculating the change in $CO_{2}$ mass fraction with time via post-processing
\end{itemize}

\subsection*{Spinodal decomposition in a binary A-B alloy}

As a second example, let us consider a two-dimensional simulation of the Cahn-Hilliard equation \cite{cahn1958free} for an A-B alloy. During spinodal decomposition, when a homogeneous binary alloy is rapidly cooled from a given temperature, the resulting domain consists of a fine-grained structure of two phases, and over time, the fine-grained structure coarsens at the expense of smaller particles. The development of a fine-grained structure from a homogeneous state is referred to as spinodal decomposition, while the coarsening mechanism is often defined as Ostwald ripening. 

\begin{figure*}[ht]
\centering
\includegraphics[width=1.0                                                                              \textwidth]{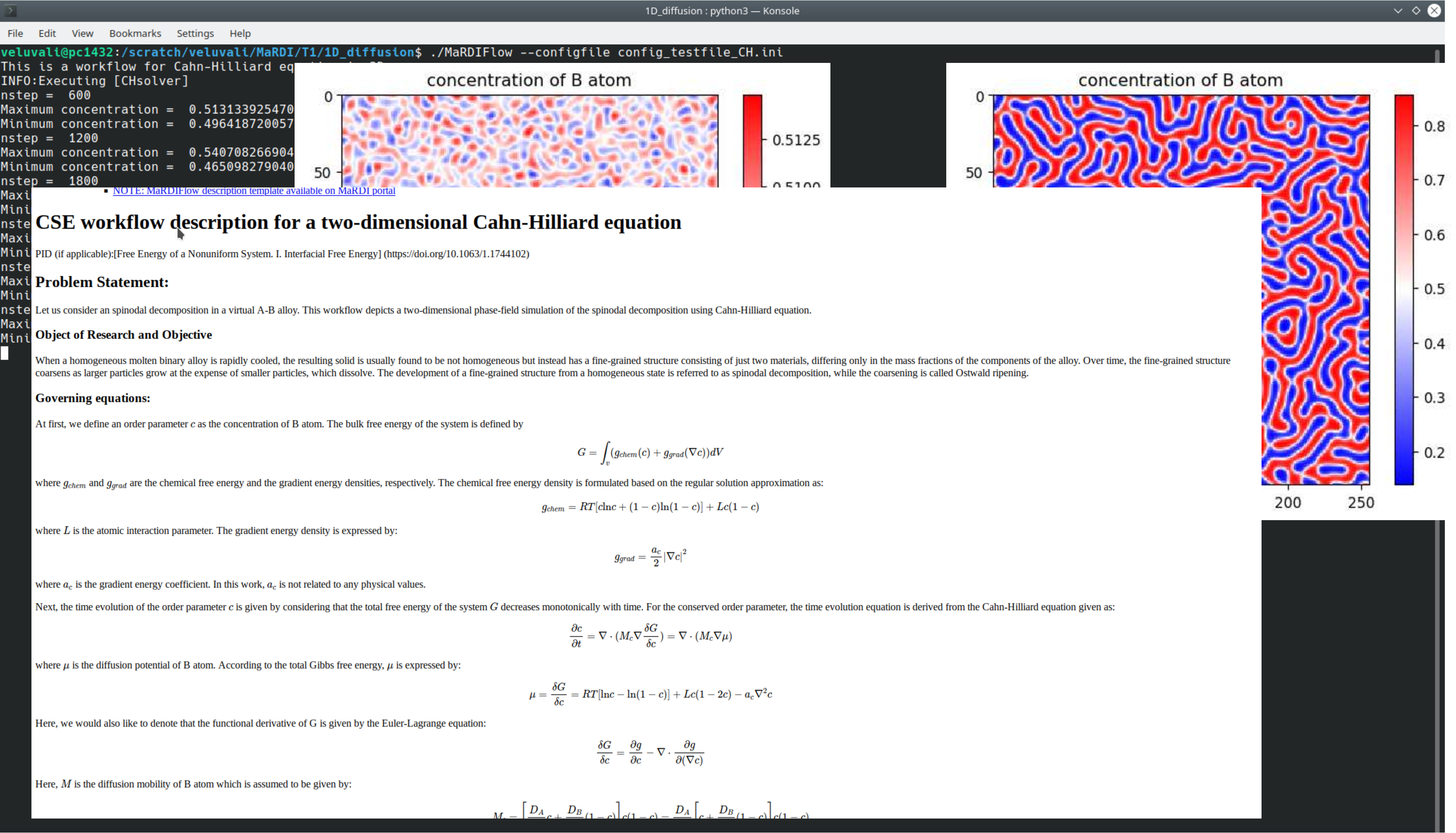}
\caption{A screenshot illustrating the workflow description of a Cahn-Hilliard Model using the \mf~tool. Screenshots of the console and simulation images as the outputs from the workflow are shown here. In addition, as a final output, the descriptive part of the tool shows the necessary meta-data to reproduce the use case.}
\label{ch_workflow}
\end{figure*}

The schematic representation of the workflow is provided in Fig.~\ref{ch_workflow}, and the set of governing equations required to simulate the phase-separation behavior between A-B alloy is given below. At first, we define an order parameter $c$ as the concentration of B atom, and the bulk free energy of the system is defined by
\begin{equation}
G = \int_v(g_{chem}(c) + g_{grad}(\nabla{c}))dV
\end{equation}
where $g_{chem}$ and $g_{grad}$ are the chemical free energy and the gradient energy densities, respectively. In this study, the chemical free energy density is formulated as:

\begin{equation} 
g_{chem} = RT[{c}\ln{c} + {1-c}\ln(1-c)] + Lc(1-c)
\end{equation}
where $L$ is the atomic interaction parameter, and the gradient energy density is given as:
\begin{equation} 
g_{grad} = \frac{a_c}{2}|\nabla{c}|^2 
\end{equation}
where $a_c$ is the gradient energy coefficient. Considering the total free energy of the system decreases with time, the temporal evolution of the order parameter $c$ is given by: 

\begin{equation} 
\frac{\partial{c}}{\partial{t}} = \nabla\cdot(M_c\nabla\frac{\delta{G}}{\delta{c}}) = \nabla\cdot(M_c\nabla{\mu})
\label{CHeq}
\end{equation}
where $\mu$ is the diffusion potential of B atom. According to classical thermodynamics, $\mu$ is generally expressed as:

\begin{equation}
\mu = \frac{\delta{G}}{\delta{c}} = RT[\ln{c} - \ln(1 - c)] + L(1 - 2c) - a_c\nabla^2{c}
\end{equation}

From \ref{CHeq},  $M_c$ is the diffusion mobility of B atom, given by:

\begin{equation} 
M_c = \Biggl[\frac{D_A}{RT}c + \frac{D_B}{RT}(1-c)\Biggr]c(1-c) = \frac{D_A}{RT}\Biggl[c + \frac{D_B}{D_A}(1-c)\Biggr]c(1-c)
\end{equation}

In the above equation, the parameters $D_A$ and $D_B$ are the the diffusion coefficients of the respective A and B atoms in the system. Lastly, herein, the Cahn-Hilliard equation is discretized by simple finite difference method, 1$^{\textrm{st}}$-order Euler method is used for time-integration, and for spatial derivatives the 2$^{\textrm{nd}}$-order central finite difference method is implemented. The workflow for the present use-case is carried out as given below:
\begin{itemize}
\item Initialize the bulk free energy and initial local concentration through an inputs object JSON file.
\item The initial configuration of the simulation domain as shown in Fig.~\ref{ch_workflow}.
\item Pass the required simulation parameters to the workflow component.
\item Time evolution of local concentration as well as the phase-separation process is captured as an output through simulation images.
\item Alongside, concentration for various timesteps is collected as an output as well.
\end{itemize}

The above workflow can be performed by using \texttt{\mf~--config config\_CH\_2D.ini} in the root directory terminal, and the resulting output shall be displayed on the screen, similar to Fig.~\ref{ch_workflow}. At the end of the workflow, the phase-separated simulation screenshots along with the corresponding equilibrium concentration are collected in the user-defined output directory. 

\section*{Summary and Outlook}

The practice to perform data and software intensive tasks has been taken hold by computational workflows. Subsequently, the rapid growth in their uptake and application on computer-based experiments presents a crucial opportunity to advance the development of reproducible scientific softwares. As a part of the MaRDI consortium \cite{mardi} on research data management in mathematical sciences, in this work, we presented a novel computational workflow framework, namely, \mf, a prototype that focuses on the automation of abstracting meta-data embedded in an ontology of mathematical objects while negating the underlying execution and environment dependencies into multi-layered vertical descriptions. Additionally, the different components are characterized by their input and output relation such that they can be used interchangeably and in most cases redundantly. 

The design specification as well as the working prototype of our RDM tool was presented through different use cases. In the present version, ~\mf~acts a command-line tool such that it enables users to handle the workflow components as abstract objects described by input to output behavior. At its core, \mf~ensures that the output generated is detailed, and a comprehensive description facilitates the reproduction of computational experiments. At first we illustrated the conversion rates of CO2 using a methanization reactor model, and later, we demonstrated the two-dimensional spinodal decomposition of a virtual A-B alloy using the Cahn-Hilliard model. Our RDM tool adheres to FAIR principles, such that the abstracted workflow components are Findable, Accessible, Interoprable and Reusable. Overall, the ongoing development of \mf~aims at covering heterogeneous use cases and act as a scientific tool in the field of mathematical sciences. 

Apart from this, we are also working towards developing an Electronic Lab Notebook (ELN) in order to visualize as well as execute the \mf~tool. The ELN will provide researchers with a user-friendly interface to interact with the tool efficiently and seamlessly. Lastly, although the present manuscript introduces our RDM tool as a working proof of concept, we plan to publish a detailed manuscript with technical details and use cases in the near future. 

\section*{Acknowledgments}%
\addcontentsline{toc}{section}{Acknowledgments}
The authors are supported by \emph{NFDI4Cat} and \emph{MaRDI}, funded by the
Deutsche Forschungsgemeinschaft (DFG), project 441926934 ``NFDI4Cat – NFDI für
Wissenschaften mit Bezug zur Katalyse'' and project 460135501 ``MaRDI – Mathematische Forschungsdateninitiative''.

\section*{Data Availability}
Results presented in this work are apart of an ongoing investigation, however a working prototype with the second use-case is available and documented at \href{https://doi.org/10.5281/zenodo.10608764}{https://doi.org/10.5281/zenodo.10608764}

\addcontentsline{toc}{section}{References}
\bibliographystyle{alphaurl}
\bibliography{mardi-flow-reactor}
  
\end{document}